\documentclass[aps,prl,twocolumn,groupedaddress]{revtex4-1}
\usepackage{graphicx}
\usepackage{times}
\usepackage{amsmath}
\usepackage{textcomp}
\newcommand{\boldnabla}{\mbox{\boldmath$\nabla$}}
\begin{document}

\title{Nonlocal Cooper pair Splitting in a \textsl{p}S\textsl{n} Junction}
\author{M. Veldhorst}
\author{A. Brinkman}
\affiliation{Faculty of Science and Technology and MESA+ Institute for Nanotechnology, University of Twente, 7500 AE Enschede, The Netherlands}

\date{\today}

\begin{abstract}
Perfect Cooper pair splitting is proposed, based on crossed Andreev reflection (CAR) in a \textsl{p}-type semiconductor-superconductor-\textsl{n}-type semiconductor (\textsl{p}S\textsl{n}) junction. The ideal splitting is caused by the energy filtering that is enforced by the bandstructure of the electrodes. The \textsl{p}S\textsl{n} junction is modeled by the Bogoliubov-de Gennes equations and an extension of the Blonder-Tinkham-Klapwijk theory beyond the Andreev approximation. Despite a large momentum mismatch, the CAR current is predicted to be large. The proposed straightforward experimental design and the 100\% degree of pureness of the nonlocal current open the way to \textsl{p}S\textsl{n}- structures as high quality sources of entanglement.
\end{abstract}
\pacs{74.45.+c, 03.65.Ud, 74.78.Na}
\maketitle

Spatially separated entangled electron pairs arise in hybrid normal-metal-superconductor structures. Andreev reflection (AR) is the conversion of an electron into a hole at the interface between a normal-metal or semiconductor and a superconductor \cite{Andreev1964}. In nonlocal (or crossed) Andreev reflection (CAR) the conversion is over two electrodes maintaining the singlet state \cite{Byers1995, Deutscher2000}. This makes CAR a promising source of locally separated entangled electrons and building block for solid state Bell inequality experiments, quantum computation and quantum teleportation \cite{Recher2001, Lesovik2002}.

The Bell inequality can only be violated when the CAR fraction is larger than 1/$\sqrt{2}$ of the total current \cite{Guhne2009}. Despite intensive investigation, the CAR fraction is usually small due to competing processes \cite{Brinkman2006, Hofstetter2009}. Aside from local AR, tunneling of a particle from one lead to another can occur. Elastic cotunneling (EC) is to lowest order in tunneling amplitude equal in magnitude and opposite in sign to CAR, resulting in a vanishing nonlocal conductance \cite{Falci2001}. Including higher order terms, which become important in more transparent junctions, unfortunately provides EC to be the dominant process \cite{Kalenkov2007}. 

\begin{figure}[b]
	\centering
		\includegraphics[width=0.47\textwidth]{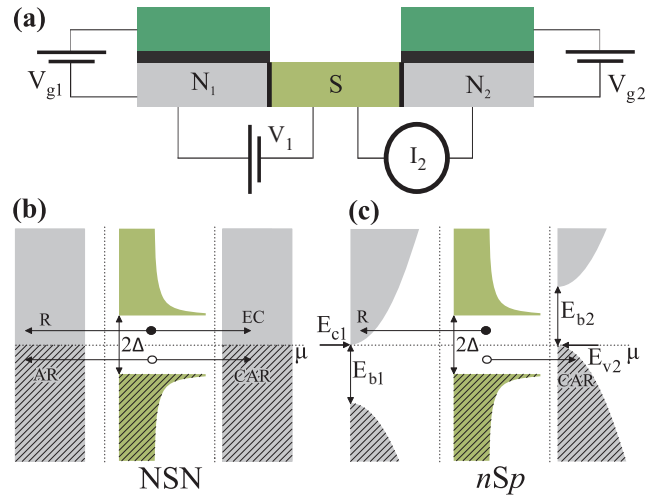}
		\caption{ (a) NSN structure with gates attached to the semiconductor. By applying gate-voltages, the semiconductor bands in each electrode can be tuned with respect to the Fermi energy. A bias voltage applied at the left NS interface results in a nonlocal conductance at the second interface. (b) In standard NSN structures an incident electron can result in normal reflection (R), AR, EC, and CAR. (c) In an $\textit{n}$S$\textit{p}$ junction it is possible to have perfect nonlocal Cooper pair splitting.}
		\label{fig:psnjunction}
\end{figure}

Several proposals have been put forward to enhance the CAR current. Using ferromagnetic-halfmetals (F) as leads can result in dominant CAR in an antiparallel magnetization alignment \cite{Deutscher2000}, though spin entanglement is then questionable. Pure nonlocal Cooper pair splitting is predicted in a superconductor-topological insulator structure \cite{Nilsson2008}, but the fabrication will be challenging. Both the electromagnetic environment \cite{Yeyati2007} and a change in the density of states (DOS) of a superconductor due to an ac bias \cite{Golubev2009} can result in dominant CAR, though the influences are expected to be small. A larger effect is predicted by Cayssol \cite{Cayssol2008} in an \textit{n}-type graphene-superconductor-\textit{p}-type graphene junction. However, full cancellation of EC and AR is only at precise biasing to the Dirac point and at a small range of the energy spectrum so that the current is not completely carried by CAR. Optimization of CAR may also be realized by using the Coulomb interaction \cite{Recher2001}. Recent experiments indicated great potential of using the energy to discriminate CAR \cite{Hofstetter2009}, although the splitting efficiency is yet small.  

Here, we propose a strategy for ideal 100\% nonlocal Cooper pair splitting, with no contributions from AR and EC, in a relatively straightforward device. Making use of the energy difference between the incoming electron and the Andreev reflected hole, in combination with bandstructure imposed forbidden energies, enables the cancellation of both AR and EC individually, while still having a significant CAR probability. This idea is shown in Fig.~\ref{fig:psnjunction}. The asymmetry of the bandstructure allows only CAR to occur since particles due to EC and AR will end up in forbidden states in the band structure, the band alignment being tunable by proper gating. 

A wide range of materials are suitable as electrodes. Examples are nanowires, where the bandgap can be tuned by the length of the wire, minigap semiconductors such as bilayer graphene, and narrow band semiconductors in general. Impurity bands can be used when there is significant density of states at an energy $\pm\Delta$ from the semiconducting bandgap, generally in the case of low doping concentrations. Energy bands or levels that arise from quantum confinement in general can be used, as long as gapped energy regions exist that prohibit AR and EC. The proposed type of energy filtering is not only of use as Andreev entangler, but can also serve as energy beam splitter in, for example, FSF devices. This opens up an alternative route towards Bell inequality experiments; the spin may be employed to split the Cooper pairs, since the energy can be utilized for the read out. 

We model the NSN system by extending the classical Blonder-Tinkham-Klapwijk model \cite{Blonder1982} to three dimensions and by including the second interface. The BTK model has been used in modeling SNS and FSF structures where the dimension of the sandwiched layer is close to $\xi$ \cite{Schussler1993}. In the present case of a $p$S$n$ junction, the need to model in three dimensions stems from the large Fermi momentum mismatch between a semiconductor and a superconductor. Because of the conservation of momentum parallel to the interfaces, a critical angle between momentum and interface normal exists above which no transfer can take place. Each of the AR, CAR and EC probabilities is characterized by an energy and angle dependent effective barrier. Since excitation energies are comparable to or larger than the Fermi energy of the semiconductor in our system, we will go beyond the Andreev approximation that takes all momenta equal. 

We describe the $p$S$n$ structure shown in Fig.~\ref{fig:psnjunction} with the time independent Bogoliubov-de Gennes equations given by
\begin{equation}
\left(
\begin{array}{cc}
\hat{H}(\textbf{r}) & \Delta(\textbf{r}) \\ 
\Delta^*(\textbf{r}) & -\hat{H}(\textbf{r}) \end{array} \right) 
\Psi(\textbf{r})=E\Psi(\textbf{r}).
\label{eq:BdG}
\end{equation} 
$\Psi(\textbf{r})=\left(u,v \right)^T$ is the wavefunction in Nambu (electron-hole) space. We assume that $\Delta(\textbf{r})=0$ in the normal regions $N_1$ ($z<0$) and $N_2$  ($z>d$), and $\Delta(\textbf{r})=\Delta_0$ in the superconducting region $S$ ($0<z<d$), $d$ being the superconductor width. The use of these rigid boundary conditions is warranted by the large Fermi momentum mismatch across the interfaces, which effectively reduces the coupling between the layers. A specular barrier is included at the interfaces, resulting in $U(\textbf{r})=H_1\delta(z)+H_2\delta(d-z)$. Inside a region we assume an isotropic bandstructure. Incorporating the conserved momentum component parallel to the interfaces into the Hamiltonian allows us to simplify Eq. (\ref{eq:BdG}) to a 1D system with an effective 1D Hamiltonian, given by $\hat{H}(z)=-\frac{\partial}{\partial z}  \frac{\displaystyle{
\hbar^2}}
{\displaystyle{
2 m^\ast(z)}}
\frac{\partial}{\partial z}+U(z)-\mu^*, $
where $\mu^*=(\mu-E_p) \cos^2\theta_\pm \mp \sqrt{E^2-\Delta^2} \sin^2\theta_\pm$. The angle $\theta_{\pm}$ is the angle between the direction of the electron (+) or hole (-) and the normal of the interface and can be found through the Snell-Descartes law $\sin{\theta_{t}}=r_k\sin{\theta_{i}}$ where $\theta_{i}$ and $\theta_{t}$ are the respective incidence and transmission angle and $r_k$ is the ratio of the incoming and transmitted moment. For large $r_k$, $\theta_t\approx0$ and particles in the superconductor travel normal to the interface. $\mu$ is the chemical potential and $E_p$ is the potential energy in each layer tuned by the gate voltages. Our ansatz for $\Psi$ in the regions $N_1$, $S$, and $N_2$, then becomes
\begin{eqnarray}
&& \Psi_{N_1}=\left(\begin{array}{cc} e^{iq_1^+z}+ b \ e^{-iq_1^+z}\\ a \ e^{iq_1^-z} \end{array}\right), 
\Psi_{N_2}= \left(\begin{array}{ll} c \ e^{iq_2^+z}\\ d \  e^{-iq_2^-z} \end{array}\right),
\nonumber
 \\
&& \Psi_S= \psi^+ \left( \begin{array} {ll} u_0\\v_0\end{array} \right)+ \psi^- \left( \begin{array} {ll} v_0\\u_0\end{array} \right),
\label{eq:Ansatz}
\end{eqnarray}
where $\psi^+$=$\alpha \ e^{ik^+z}+\chi \ e^{-ik^+z}$, $\psi^-$=$\beta \ e^{-ik^-z} + \eta \ e^{ik^-z}$ and $u_0^2=1-v_0^2=\frac{1}{2}(1+\frac{\sqrt{E^2-\Delta^2}}{E})$ \cite{Blonder1982}. Using conservation of $k^{||}$ we find the moments in the $z$-direction in each layer given by $k^{\pm}=\cos\theta_{\pm} \sqrt{\frac{2m^*}{\hbar^2} \left(\mu - E_p \pm \sqrt{E^2-\Delta^2}\right)}$. This implies a critical angle given by $\theta_C(E)=\arcsin (r_k^{-1})$. In a $\textit{p}$S$\textit{n}$ junction with $E_p \neq 0$ in the electrodes, CAR is enhanced since CAR has the lowest critical angle as compared to the other scatter processes \cite{Cayssol2008}. 

The system can be solved by applying the boundary conditions $\Psi_{N_{1,2}}(0^-,d^+)=\Psi_S(0^+,d^-)$ together with $\frac{\hbar^2}{2m^*_S}\frac{\partial\Psi_S}{\partial z}(0^+,d^-) -\frac{\hbar^2}{2m^*_{N_{1,2}}}\frac{\partial\Psi_{N_{1,2}}}{\partial z}(0^-,d^+) =\pm H_{1,2} \Psi$. This extended BTK model reproduces the results in NSN structures found previously by other models. The nonlocal conductance $G_{\text{NL}}$ vanishes due to the cancellation of CAR by EC in the tunnel limit \cite{Falci2001}, while EC is dominant in transparent regimes \cite{Kalenkov2007}, and the electrode separation distance dependence of $G_{\text{NL}}$ is exponential. Charge imbalance is not taken into account in this model, but the CAR enhancement effects as described in this Letter also occur in the range $E<\Delta$ at temperatures $T \ll T_c$, where this effect is absent \cite{Pethick1980}.

In order to enhance the CAR current to 100\%, $p$ and $n$ type semiconductors will be implemented now as the two electrodes. We investigate the generic example of a semiconductor possessing a band gap $E_b\gg \Delta$, one valence band and one conduction band, each with parabolic dispersion. Even scattering processes at energies within the forbidden semiconducting band gap have nonzero probabilities at the interfaces, influencing other processes, and thus need to be taken into account when solving the wave equations. Still, the wave function of these particles decay exponentially in the electrodes and do not contribute to the current by themselves, since we consider electrode lengths much larger than $\xi$.

\begin{figure}
	\centering
		\includegraphics[height=0.29\textwidth]{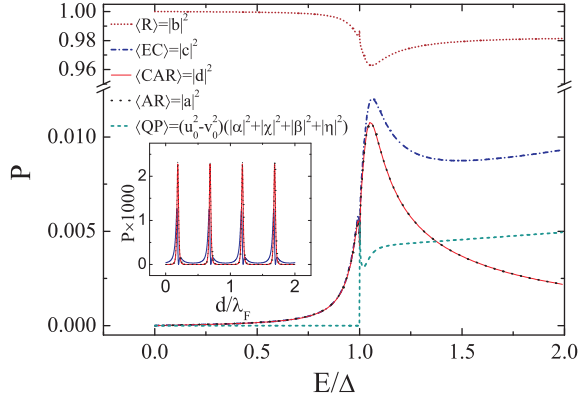}
			\caption{Energy dependence of the averaged reflection and transmission probabilities. $Z\!=\!0$ and $E_{c1}\!=\!\mu\!=\!E_{v2}=$ 5.8 $\times$ 10$^4\Delta$, resulting in an effective energy dependent barrier. Consequently, the probability $\left<CAR\right>$ is significant and almost equal to $\left<EC\right>$ for $E\!\!<\!\!\Delta$. Because of the small superconducting width $d\!=\!\frac{1}{2}\xi$, $\left<AR\right>$ does not reach unity and $\left<R\right>$ does not vanish at $E\!\!=\!\!\Delta$. The inset shows the dependence of the probabilities on the width of the superconductor at $E\!=\!\frac{1}{2} \Delta$; Fabry-Perot resonances occur on the Fermi wavelength scale.}
		\label{fig:scatterprobabilities}
\end{figure}

Figure \ref{fig:scatterprobabilities} shows the numerically obtained probabilities for perpendicular incidence, zero bias, transparent interfaces (the BTK barrier strength $Z=\frac{H}{\hbar v_F}=0$ \cite{Blonder1982}) and in a regime of large momentum mismatch. The probabilities are given by the absolute squared values of the prefactors in Ansatz Eq.~(\ref{eq:Ansatz}). Normal reflection (R), EC, AR, and CAR probabilities are considered at the respective interfaces, while the quasiparticle (QP) states in the superconductor are taken far from the interface thereby vanishing at energies below $\Delta$. In the regime of large momentum mismatch, the angle dependence below the critical angle and the bias and gating dependence of the probabilities follow $P(E)=P_0(E) | \frac{E-eV-E_{c,1}}{E} |\cos^2(\theta)$, where $P_0$ refers to the unbiased probability at perpendicular incidence. As an example, we consider Al as the superconductor, resulting in a large effective barrier originating from a large ratio $\mu/\Delta=5.8 \times 10^4$. Despite the effective barrier, CAR is found to have a considerable magnitude for $d$ being close to the Bardeen-Cooper-Schrieffer coherence length $\xi=\frac{\hbar v_{F,S}}{\pi\Delta}$. Averaging the probabilities by $\left<P\right>=\int^{d+1/2\lambda_{F,S}}_{d-1/2\lambda_{F,S}}P(z)dz$ with $\lambda_{F,S}$ the Fermi wavelength in the superconductor is necessary in order to let Fabry-Perot resonances vanish, in practice washed out by roughness.  

The current density in the electrodes is obtained from
\begin{equation}
J_d^z=\frac{1}{V_d} \sum_{\textbf{k},\sigma,\pm} \textbf{J}_q(\textbf{k},\pm)\hat{\textbf{e}}_zf(\textbf{k},\pm).
\nonumber
\end{equation}
Here, $d$ is the dimension and $V_d$ the volume of the system, the sum $\pm$ is over electrons and holes, $f$ is the nonequilibrium distribution and the charge current is defined by $\textbf{J}_{q,e(h)}=\frac{e\hbar}{m}\textrm{Im}\left[u^*(v^*)\boldnabla u(v)\right].$
From this it follows that the CAR and EC current are opposite in sign, since the respective group velocities in $N_2$, given in the ansatz Eq.~(\ref{eq:Ansatz}), are opposite in sign. Finally, the current in an electrode becomes $I=\frac{A}{2\pi^3\hbar^2} \int dE f \int_0^\frac{\pi}{2} d\theta  \cos{\theta} \sin{\theta} \sum_{\pm} |\textbf{k}|mJ_q^z$, where $A$ is the cross-sectional area. The integration is over all energy modes that contribute to the tunneling, limited by the lowest DOS of the initial and final state for a certain process. Even nonideal $p$S$n$ junctions with nonvanishing DOS in the bandgap or improper Fermi level aligning enhance CAR, since the DOS will be lowest for the AR and EC processes. The distribution functions are given by $f_{1}=f_0(E-eV_{N1})-f_0(E-eV_S)$ and $f_{2,e(h)}\!=\!f_0(E-eV_{N1})-f_0(E-e\, \textrm{max}[V_s;V_{N2}(-V_{N2})])$, with $f_0(E)$ the Fermi distribution function. Positive or negative biasing at the second electrode decreases the EC or CAR processes, respectively. Current flowing to the superconductor is defined positive, so AR and CAR are positive in sign and EC negative.

\begin{figure}
	\centering
		\includegraphics[height=0.29\textwidth]{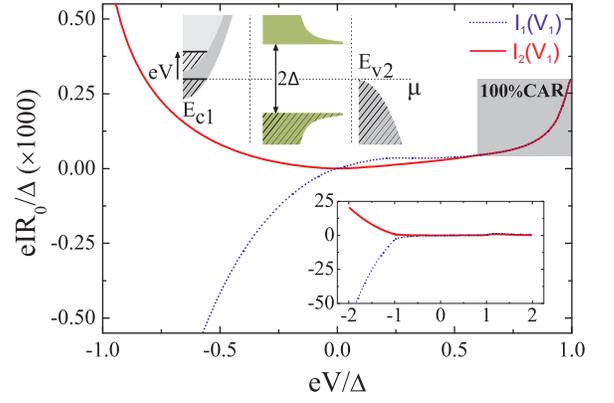}
			\caption{ Local ($I_1$) and nonlocal ($I_2$) current dependence on the bias voltage ($V_1$) for a fixed gate voltage $eV_{g1}\!=\!0.6\Delta$ (see upper inset). $E_{c1}+0.6\Delta\!=\!\mu\!=\!E_{v2}=$ 5.8 $\times$ 10$^4\Delta$, $d\!=\!\frac{1}{2}\xi$ and $Z\!=\!0$. The momentum mismatch results in an effective barrier. $R_0$ is the Sharvin resistance at $eV\!=\!\Delta$. Negative biasing ($eV\!\!<\!\!0$) leads to a nonlocal current due to EC, whereas positive biasing ($eV\!\!>\!\!0$) results in CAR. AR is possible up to 0.6$\Delta$, so that in the range 0.6$\Delta\!\!<\!\!eV\!\!<\!\!\Delta$ perfect Cooper pair splitting occurs. QP current appears at $eV\!\!>\!\!\Delta$ shown in the lower inset.}
		\label{fig:fixedgate}
\end{figure}
 
Figure \ref{fig:fixedgate} shows the $IV$ characteristics for a $p$S$n$ junction with fixed gate voltages, so that $E_{c1}+0.6\Delta=\mu=E_{v2}$. At negative bias across the first interface, the first electrode has available states above and below $\mu$ whereas the second electrode has only states below $\mu$. AR and EC are therefore possible, while CAR is prohibited. For positive bias voltages, direct electron transfer is no longer possible and the nonlocal current is carried by CAR only. AR is significantly reduced by the critical angle and limited DOS and totally vanishes above eV$>$0.6$\Delta$, where available states for AR are absent. The device works as a perfect Cooper pair splitter for 0.6$\Delta$$<$eV$<$$\Delta$. Above the gap a quasiparticle current appears in the superconductor resulting in a lower splitting efficiency.

\begin{figure}
     \begin{center}
\includegraphics[width=0.42\textwidth]{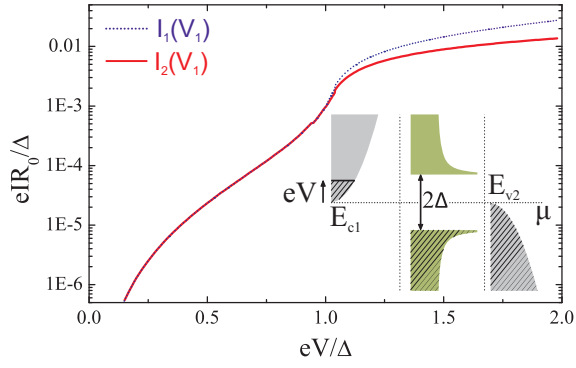}
    \end{center}
\caption{Local ($I_1$) and nonlocal ($I_2$) current dependence on the bias voltage ($V_2$) for ideal gating, $E_{c1}\!=\!\mu\!=\!E_{v2}=$ 5.8 $\times$ 10$^4\Delta$ regardless of the bias voltage. $d\!=\!\frac{1}{2}\xi$ and $Z\!=\!0$. The momentum mismatch results in an effective barrier. $R_0$ is the Sharvin resistance at $eV\!=\!\Delta$. For $eV\!\!<\!\!0$ no process is possible, but in the range $0\!\!<\!\!eV\!\!<\!\!\Delta$ only CAR is possible and we observe pure entangled current $I_1\!=\!I_{2}$. QP current appears at $eV\!\!>\!\!\Delta$, lowering the CAR fraction.}
\label{fig:idealgate}
\end{figure}

Maximizing the bias regime in which CAR = 100\% can be achieved by tuning the gate voltage such that always $E_{c1}=\mu=E_{v2}$, irrespective of the bias voltage. This bias situation is shown in Fig.~\ref{fig:idealgate}. For all positive bias voltages below $\Delta$, CAR = 100\%. AR and EC are forbidden due to the bandgap in the electrodes. The DOS for incoming electrons is equal to the DOS for outgoing holes and there is no critical angle lowering the CAR probability, since $E_P=0$. Consequently, the nonlocal current is maximized to a relatively large value, being typically a few $\hbox{\textmu}$A.

We now address the question what materials would be best suited. The discrimination between AR, EC, and CAR leading to 100\% crossed Andreev reflection in a $\textit{p}$S$\textit{n}$ junction is caused by: the forbidden band gap (which is our main effect leading to 100\% CAR), the variation in density of states and the critical angle. The elasticity mandatory for these three effects is typically a less stringent condition than the superconductor width being comparable to $\xi$, necessary to have a significant nonlocal current. With current nanolithography methods and using Al as superconductor these requirements are easily fullfilled. For the semiconductors, InAs two-dimensional electron gases are ideal candidates, since no Schottky barrier is formed in contact to Al \cite{Schapers1998}. The effect of the critical angle may vanish in the diffusive limit, but perfect Cooper pair splitting due to the forbidden bandgap remains robust against disorder. Nb/InAs structures are therefore also suited. Even though $\xi_{Nb}<\xi_{Al}$, the larger superconducting gap of Nb simplifies the band alignment and increases the magnitude of the nonlocal current. Schottky barriers reduce the nonlocal current, but the splitting is found to remain ideal when using a nonzero barrier strength. Al/GaAs heterostructures are, therefore, suitable as well \cite{Taboryski1996}. Finally, we mention that electronic gate controllable InAs nanowires have been contacted to Al with high interface transparency \cite{Doh2005}, making it an ideal system for entanglement experiments where a reduced number of propagating modes are required.

In conclusion, we have proposed a $\textit{p}$S$\textit{n}$ junction that can be used to prepare a pure Bell state by forward biasing and can act as a perfect Cooper pair splitter by reversed biasing, while having significant currents.

Discussions with A. A. Golubov, B. C. Kaas, C. J. M. Verwijs and H. Hilgenkamp are gratefully acknowledged. This work is supported by the Netherlands Organization for Scientific Research (NWO).


\begin{thebibliography}{33}
\bibitem{Andreev1964} A.F. Andreev, Zh. Eksp. Teor. Fiz. \textbf{46}, 1823 (1964) [Sov. Phys. JETP \textbf{19}, 1228 (1964)].
\bibitem{Byers1995} J.M. Byers and M.E. Flatt\'e, Phys. Rev. Lett. \textbf{74}, 306 (1995).
\bibitem{Deutscher2000} G. Deutscher and D. Feinberg, Appl. Phys. Lett. \textbf{76}, 487 (2000).
\bibitem{Recher2001} P. Recher, E.V. Sukhorukov, and D. Loss, Phys. Rev. B \textbf{63}, 165314 (2001).
\bibitem{Lesovik2002} G.B. Lesovik, T. Martin and G. Blatter, Eur. Phys. J B \textbf{24}, 287 (2001); N.M. Chtchelkatchev \textit{et al.}, Phys. Rev. B \textbf{66}, 161320(R) (2002). N. M. Chtchelkatchev, JETP Lett. \textbf{78}, 230 (2003) [Pis'ma Zh. Eksp. Teor. Fiz. \textbf{78}, 265 (2003)]. K. V. Bayandin, G.B. Lesovik, and T. Martin, Phys. Rev B \textbf{74}, 085326 (2006). S. Kawabata, J. Phys. Soc. Jap. \textbf{70}, 1210 (2001).
\bibitem{Guhne2009} O. G\"uhne and G. T$\acute{\textrm{o}}$th, Phys. Rep. \textbf{474}, 1 (2009).
\bibitem{Brinkman2006} A. Brinkman, A.A. Golubov, Phys. Rev. B \textbf{74}, 214512 (2006). S. Russo \textit{et al.} Phys. Rev. Lett. \textbf{95}, 027002 (2005). P. Cadden-Zimansky and V. Chandrasekhar, Phys. Rev. Lett. \textbf{97}, 237003 (2006). A. Kleine \textit{et al.}, Europhys. Lett. \textbf{87}, 27011 (2009). D. Beckmann, H.B. Weber and H.v. L\"ohneysen, Phys. Rev. Lett. \textbf{93}, 197003 (2004).
\bibitem{Hofstetter2009} L. Hofstetter \textit{et al.}, Nature \textbf{461}, 960 (2009). L. G. Herrmann \textit{et al.}, Phys. Rev. Lett. \textbf{104}, 026801 (2010).
\bibitem{Falci2001} G. Falci, D. Feinberg, F.W.J. Hekking, Europhys. Lett. \textbf{54}, 255 (2001).
\bibitem{Kalenkov2007} M.S. Kalenkov, A.D. Zaikin, Phys. Rev. B \textbf{75}, 172503 (2007).
\bibitem{Nilsson2008} J. Nilsson, A.R. Akhmerov and C.W.J. Beenakker, Phys. Rev. Lett. \textbf{101}, 120403 (2008).
\bibitem{Yeyati2007} A. Levy Yeyati \textit{et al.}, Nature Phys. \textbf{3}, 455 (2007).
\bibitem{Golubev2009} D.S. Golubev, A.D. Zaikin, Europhys. Lett. \textbf{86}, 37009 (2009).
\bibitem{Cayssol2008} J. Cayssol, Phys. Rev. Lett. \textbf{100}, 147001 (2008).
\bibitem{Blonder1982} G.E. Blonder, M. Tinkham and T.M. Klapwijk, Phys. Rev. B \textbf{25}, 4515 (1982).
\bibitem{Schussler1993} U. Sch$\ddot{\textrm{u}}$ssler and R. K$\ddot{\textrm{u}}$mmel, Phys. Rev. B \textbf{47}, 2754 (1993). T. Yamashita, S. Takahashi and S. Maekawa, Phys. Rev. B \textbf{68}, 174504 (2003). 
\bibitem{Pethick1980} C.J. Pethick and H. Smith, J. Phys. C: Solid State Phys. \textbf{13}, 6313 (1980).
\bibitem{Schapers1998} H. Takayanagi, T. Akazaki, J. Nitta, Phys. Rev. Lett. \textbf{75}, 3533 (1995); Th. Sch\"apers \textit{et al.}, Appl. Phys. Lett. \textbf{73}, 2348 (1998).
\bibitem{Taboryski1996} R. Taboryski \textit{et al.}, Appl. Phys. Lett. \textbf{69}, 656 (1996)
\bibitem{Doh2005} Y.J. Doh \textit{et al.}, Science \textbf{309}, 272 (2005)
\end{thebibliography}
\end{document}